# Strongly coupled edge states in a graphene quantum Hall interferometer

Thomas Werkmeister[1*], James R. Ehrets[2], Yuval Ronen[2,3], Marie E. Wesson[1], Danial Najafabadi[4], Zezhu Wei[5,6], Kenji Watanabe[7], Takashi Taniguchi[8], D.E. Feldman[5,6], Bertrand I. Halperin[2], Amir Yacoby[1,2], Philip Kim[1,2 *]

[1]John A. Paulson School of Engineering and Applied Sciences, Harvard University, Cambridge, MA 02138, USA
[2]Department of Physics, Harvard University, Cambridge, MA 02138, USA
[3]Department of Condensed Matter Physics, Weizmann Institute of Science, Rehovot 7610001, Israel
[4]Center for Nanoscale Systems, Harvard University, Cambridge, MA 02138, USA
[5]Department of Physics, Brown University, Providence, Rhode Island 02912, USA
[6]Brown Theoretical Physics Center, Brown University, Providence, Rhode Island 02912, USA
[7]Research Center for Functional Materials, National Institute for Materials Science, 1-1 Namiki, Tsukuba 305-0044, Japan
[8]International Center for Materials Nanoarchitectonics, National Institute for Materials Science, 1-1 Namiki, Tsukuba 305-0044, Japan
*e-mail: werkmeister@g.harvard.edu;  pkim@physics.harvard.edu

**Electronic interferometers using the chiral, one-dimensional (1D) edge channels of the quantum Hall effect (QHE) can demonstrate a wealth of fundamental phenomena. The recent observation of phase jumps in a Fabry-Pérot (FP) interferometer revealed anyonic quasiparticle exchange statistics in the fractional QHE. When multiple integer edge channels are involved, FP interferometers have exhibited anomalous Aharonov-Bohm (AB) interference frequency doubling, suggesting putative pairing of electrons into 2$e$ quasiparticles. Here, we use a highly tunable graphene-based QHE FP interferometer to observe the connection between interference phase jumps and AB frequency doubling, unveiling how strong repulsive interaction between edge channels leads to the apparent pairing phenomena. By tuning electron density in-situ from filling factor $\nu < 2$ to $\nu > 7$, we tune the interaction strength and observe periodic interference phase jumps leading to AB frequency doubling. Our observations demonstrate that the combination of repulsive interaction between the spin-split $\nu = 2$ edge channels and charge quantization is sufficient to explain the frequency doubling, through a near-perfect charge screening between the localized and extended edge channels. Our results show that interferometers are sensitive probes of microscopic interactions and enable future experiments studying correlated electrons in 1D channels using our highly tunable platform.**

Electrons in 1D quantum systems exhibit striking phenomena, including the breakdown of Fermi liquid theory and quasiparticle formation in favor of collective modes[1]. Likewise, electrons confined to two dimensions and subjected to perpendicular magnetic fields exhibit the quantum Hall effects (QHEs)[2]. Although the microscopic details of QHE states are still an active area of research[3,4], their low-energy transport properties are known to be governed by chiral, 1D edge channels[5–8]. These edge channels (ECs) conduct charge ballistically, allowing for phase-coherent electronic experiments[9,10]. In particular, electronic Fabry-Pérot (FP) QHE interferometry[11–13], was performed extensively in GaAs, culminating in the observation of interference phase jumps as evidence for anyonic statistics of fractional quasiparticles[14–17]. Recently, FPs were developed in graphene, which showed Aharonov-Bohm (AB) interference of integer ECs[18–20], with oscillation periodicity set by the magnetic flux quantum for electrons $\Phi_0 \equiv h/e$. Our previous design[18] utilized graphite gates encapsulating the graphene channel, which screened bulk charges. Without such screening layers[21], however, interferometers exhibit 'Coulomb dominated' (CD) behavior in which strong coupling of the interfering EC to localized compressible states in the bulk determines the oscillation periodicity and obscures the expected AB oscillations[13,22–24].

When bulk charges were strongly screened, GaAs FPs showed unexpected doubling of the AB oscillation frequency and shot noise corresponding to charge $2e$ when interfering the outermost EC with the bulk of the interferometer in filling $2.5 \leq \nu \leq 4.5$, suggesting a possibility of 'pairing' of elementary charges[25]. Furthermore, the coherence and periodicity of the interfering outer EC were related to the coherence and the enclosed flux of the adjacent inner EC[26], and the 'pairing' phenomena only occurred when the outer two modes belonged to the same spin-split Landau level[27]. Independently, single-electron capacitance measurements in GaAs quantum dots revealed that tunneling into the edge of the dot corresponded to the entrance of two electrons rather than one for $\nu \geq 2$, and that near $\nu \approx 2.5$ the charging peaks follow doubled magnetic flux frequency[28].

Mechanisms of electron pairing are important questions in emergent phenomena, e.g. high-temperature superconductivity[29] and the $\nu = 5/2$ fractional QHE state in GaAs[30] and bilayer graphene[31,32]. However, theoretical work concerning FP interferometers was able to explain the doubled AB oscillation frequency based on a microscopic model without explicit introduction of electron pairing, though explaining other related phenomena in GaAs remains challenging[33]. In this work, we experimentally address the microscopic mechanism of coupling between QHE edges by elucidating the relation between AB oscillation phase jumps and frequency doubling, employing a highly tunable QHE FP interferometer with strongly screened bulk charge in graphene.

**Interferometer design and tuning**

We designed a graphene-based FP interferometer tuned by a local gate array (Fig. 1a). The FP cavity is defined electrostatically using separated graphite top-gates (Methods and Supplementary Information). Metal bridges contact each top-gate, and we additionally suspend metal bridges over the two quantum point contacts (QPCs), illustrated in Fig. 1b. By applying voltages $V_{QPC1}$ and $V_{QPC2}$ to these suspended bridge gates, we can tune each QPC independently while keeping the filling factor of the surrounding regions fixed.

In our experiments, we measure the diagonal conductance $G_D$, as shown in Fig. 1b. In the regime that we study, $G_D = \frac{e^2}{h} \nu_{QPC}$ where $\nu_{QPC}$ counts the number of edge channels transmitted through the device, with a partially transmitted channel counted as fraction[34,35]. To characterize the QPC transmissions, we measure $G_D$ as a function of the bottom-gate voltage and split-gate voltage for each QPC with the bulk of the interferometer tuned to $\nu = 2$ at $B = 6$ T (Supplementary Information). At $\nu = 2$, there are two spin-split Landau levels, of which the lower energy spin species hosts an EC closer to the effective boundary of the sample. Hence, we refer to the EC belonging to the lower (higher) energy spin species as the 'outer' ('inner') EC. Once appropriate bottom-gate and split-gate voltages are set, we tune $V_{QPC1}$ and $V_{QPC2}$, voltages applied on the suspended bridges to control the individual QPC transmissions. Fig. 1c and Fig. 1d show the measured $G_D$ as a function of $V_{QPC1}$ and $V_{QPC2}$, respectively, with the other QPC fully open. $G_D$ shows plateaus at $(0,1,2)\frac{e^2}{h}$, corresponding to (neither, outer, both) ECs transmitted. In this regime, we define $T_{QPC} \equiv G_D \frac{h}{e^2}$ as the transmission of the QPC[34].

Tuning to partial transmission of the inner EC for both QPCs, $T_{QPC1} = T_{QPC2} = 1.5$, we observe high-visibility conductance oscillations as a function of plunger gate voltage $V_{PG}$, which tunes the filling factor $\nu_{PG}$ under the plunger gate, in Fig. 1e. Similarly, we tune to $T_{QPC1} = T_{QPC2} = 0.5$ and measure conductance oscillations on the outer EC in Fig. 1f. In both cases, oscillations are largest for $\nu_{PG} < 0$, which corresponds to a fully gate-defined interference path since electrons are depleted under the gate. Increasing $\nu_{PG}$ brings the interfering edge closer to the etched graphene boundary, inducing dephasing[18]. Notably, the inner EC oscillations survive until $\nu_{PG} = 2$, when it flows close to the etched boundary of the graphene, while the outer EC reaches the boundary by $\nu_{PG} = 1$. Another difference is the apparent irregularity of the oscillations on the outer EC compared to the inner EC, which we will understand in this work.

**Phase jumps and AB oscillation frequency transition**

High-visibility oscillations allow us to probe the dependence of interference phase $\theta$ on magnetic field variation $\delta B$ and gate voltage variations, which distinguishes the AB from the CD regimes[13,15,18,19,22]. For small variations in field and gate voltages in the AB regime, we expect $\delta\theta/2\pi \approx A\delta B/\Phi_0 + C_{PG}\delta V_{PG}/e + C_{MG}\delta V_{MG}/e$, where $A$, $C_{PG}$, and $C_{MG}$, are the (approximately constant) area enclosed by the interfering EC, interfering EC – plunger gate capacitance, and interfering EC – middle gate capacitance, respectively. Importantly, $V_{MG}$ also directly tunes the electron density in the interferometer. Sweeping $V_{MG}$ over a large range with both QPCs fully open reveals Hall conductance plateaus reflecting the filling factor $\nu$ in the interferometer (Fig. 2a). Data in the remaining panels of Fig. 2 were taken with the QPCs set to $T_{QPC1} = T_{QPC2} = 0.5$. Near the lowest density of the $\nu = 2$ plateau (Fig. 2b), we observe a typical AB interference pattern. Constant phase stripes ($\delta\theta = 0$) trace out a negative slope $\delta V_{PG}/\delta B$ with magnetic field period $\Delta B$ yielding $\Phi_0/\Delta B = 1.13$ μm$^2$, matching the designed area $A = 1.16$ μm$^2$. Plunger gate period $\Delta V_{PG}$ yields $1/\Delta V_{PG} = 19.2$ V$^{-1}$. Increasing $\nu$ using $V_{MG}$ reveals more complicated interference patterns in Fig. 2c-d. Periodic shifts in the interference pattern persist and modulate until near the center of $\nu = 4$, as seen in Fig. 2e, when a simple stripe pattern returns. However, now $\Phi_0/\Delta B = 2.32$ μm$^2$ and $1/\Delta V_{PG} = 36.3$ V$^{-1}$, both approximately doubled from Fig. 2b. Since $A$ is fixed, a doubling of $\Phi_0/\Delta B$ indicates oscillations with $\Phi_0/2 = h/2e$ periodicity instead of $\Phi_0$ so that $\Phi_0/2\Delta B = 1.16$ μm$^2$. Similarly, assuming a fixed $C_{PG}$, then $1/\Delta V_{PG}$ doubling corresponds to adding twice as many electrons to the system per flux quanta. Both could be interpreted as an effective charge $e^* = 2e$ for the interfering particle, as in GaAs[25–27], but our observations indicate a different interpretation. We observe the transition to the AB frequency-doubled regime at fixed $B$ by sweeping $V_{MG}$ and observing oscillations with $V_{PG}$, as shown in Fig. 2f. Remarkably, the transition occurs continuously. From the top panel, $\Phi_0$ interference is apparent. As $V_{MG}$ increases, periodic phase jumps begin to appear. Both the $V_{MG}$ spacing and magnitude of the phase jumps increase, until eventually the most apparent periodicity corresponds to $\Phi_0/2$ oscillations (i.e., doubled frequency $2\Phi_0^{-1}$).

To better understand the phase jumps, we use a general relation between charge and phase in FP interferometers[36]. When a single EC passes through the two constrictions with weak backscattering, the interference phase seen by the device at zero temperature is $\theta = 2\pi Q + \theta_0$, mod $2\pi$, where $Q$ is the total electron charge (in units $e$) in the region between the two scattering points and $\theta_0$ is a constant for small variations in $B$, $V_{PG}$, and $V_{MG}$. In our experimental regime, $\nu \geq 2$, we expect this relation to hold with $Q = Q_1 + Q_2$, where $Q_1$ is the total charge residing in the lowest spin-split Landau level and $Q_2$ is the charge in the higher energy spin state (and also higher Landau levels). $Q_1$ can vary continuously since the outer EC is connected to the source and drain charge reservoirs. In contrast, $Q_2$ is required to be integer, as the corresponding energy levels are isolated through the incompressible QHE bulk. An integral change in $Q_2$ has no observable effect on the interference signal unless it produces a non-integral change in $Q_1$ due to Coulomb coupling between the two types of charge. Hence, we can redefine $\theta$ to include only the charge $Q_1$ in the lowest spin-split Landau level, and the values $Q_1$ in the ground state of the interferometer determine $\theta$. Following similar models used to understand the CD regime[15,24,37] and considering small changes in $Q_1$ and $Q_2$, we expand the change in ground state energy $E = K_1 \delta Q_1^2 + K_2 \delta Q_2^2 + 2K_{12} \delta Q_1 \delta Q_2$, where $K_i$ is the charging energy of the charge species $i$ and $K_{12}$ describes the mutual capacitive coupling between them. Energetic stability requires that $|K_{12}|^2 \leq K_1 K_2$. Within this capacitive coupling model, when $Q_2$ increases by 1, the charge $Q_1$ correspondingly decreases by a discrete (generically non-integral) amount $\Delta Q_1$ to screen the added charge, leading to a phase shift $\Delta \theta/2\pi = \Delta Q_1 = -K_{12}/K_1$.

By taking 1D fast Fourier transforms (FFTs) along lines parallel to the phase jumps[14,15], we extract several values of $\Delta\theta/2\pi$ near the center of the periodicity transition in Fig. 3a. We observe that the locations where the phase jumps occur (marked in Fig. 3b) follow a steeper slope than the slope $\delta V_{PG}/\delta V_{MG}$ of constant phase lines of the main interference oscillation in the $V_{MG}$-$V_{PG}$ planes. A steeper slope also occurs in the $B$-$V_{PG}$ plane (Fig. 2c-d). Moreover, these phase jump lines have negative slopes $\delta V_{PG}/\delta B < 0$, like the constant phase lines of AB oscillations. This observation is in sharp contrast to the phase jumps reported in the FP interferometer operated in the fractional QHE regime[14,15] or in the FP interferometer operated in the integer CD regime[37], where phase jump lines follow positive slope $\delta V_{PG}/\delta B > 0$. The different slope suggests a different structure to the energy levels that are being populated in our sample. Considering that the outer EC is partitioned at the QPCs, while the inner ECs are well isolated, we hypothesize that the charging events seen as phase jumps represent charge added to the annular, closed inner EC, illustrated in Fig. 3. The dominant coupling $K_{12}$ is directly between the outer and inner $\nu = 2$ ECs. Any charges added to higher Landau levels or to localized states in the bulk are not measurably coupled to the outer EC, presumably because of effective screening by the gates.

**AB frequency doubling from strongly coupled QHE edge states**

We provide further evidence for capacitively coupled QHE edges tuning the AB frequency in Fig. 4. At fixed $V_{MG}$ in the transition regime, we compare interference in the $B$-$V_{PG}$ plane for the inner EC, Fig. 4a, to the outer EC, Fig. 4b. This direct comparison is only possible because we can control QPC transmissions independently of bulk filling. We observe that the slope of the oscillation maxima on the inner EC (dotted lines in Fig. 4a) matches the slope of the phase jump lines on the outer EC (dotted lines in Fig. 4b). Reducing the transmission for the inner EC, the interference maxima in Fig. 4a become sharper charging resonances, corresponding to charge $Q_2 \to Q_2 + 1$ through the inner EC. When the transmission of the inner EC vanishes, the inner EC is fully disconnected from the source and drain charge reservoirs, and the outer EC is now partitioned at the QPCs to form a new interference path (shown in the left panel in Fig. 4b). Since the electrostatic configurations for Fig. 4a and Fig. 4b are identical, the regions in between the phase jump lines in Fig. 4b correspond to fixed $Q_2$, and we see that the interference phase on the outer EC shifts when the charge on the inner EC discretely changes.

Taking Fourier transform of the interference signal provides further understanding of interactions between the two ECs involved in the interference. The bottom panels of Fig. 4a and 4b show the 2D FFTs of the corresponding interference patterns in in the $B$-$V_{PG}$ planes. For interference of the inner EC (Fig. 4a), we observe a simple FFT pattern of peaks corresponding to the fundamental frequency of the inner EC $\boldsymbol{f}_i$, a vector containing the peak position in the 2D FFT, and its harmonics ($n\boldsymbol{f}_i$, where $n$ is an integer). The FFT pattern of the outer EC interference (Fig. 4b) exhibits a more complicated lattice of Fourier peaks. If we label one of the dominant peaks as the fundamental frequency of the outer EC, $\boldsymbol{f}_o$, we can then identify the rest of the peaks by addition or subtraction of the same vector $\boldsymbol{f}_i$ evident in the inner EC data. The lowest order peaks correspond to the sum $\boldsymbol{f}_{o+i} = \boldsymbol{f}_o + \boldsymbol{f}_i$ and the difference $\boldsymbol{f}_{o-i} = \boldsymbol{f}_o - \boldsymbol{f}_i$. We show a similar Fourier lattice construction in Extended Data Fig. 1 for interference in the $B$-$V_{MG}$ plane.

By tuning $V_{MG}$, we modulate the filling factor of the interferometer cavity in a wide range and observe the evolution of the interference patterns and corresponding peaks for the outer (inner) EC in Extended Data Fig. 2 (3). As in Fig. 2, phase jumps appear only within the periodicity transition. Fig. 4c shows the magnitude of individual phase jumps as a function of $V_{MG}$. We find that the phase jump continuously evolves from $\Delta\theta/2\pi \approx 0$ ($V_{MG} < 0.6$ V) through the periodicity transition to $\Delta\theta/2\pi \approx -1$ ($V_{MG} > 1.6$ V), corresponding to the strongly coupled limit $K_{12}/K_1 \approx 1$. The transition regime marked by non-trivial phase jumps spans from the appearance of the inner EC ($V_{MG} \approx 0.6$ V) to the strongly coupled outer two EC limit ($V_{MG} \approx 1.6$ V).

The Fourier peaks' evolution tuned by $V_{MG}$ provides insight into the interaction between ECs. Fig. 4d displays the normalized Fourier peak intensity as a function of $V_{MG}$. The amplitude of the Fourier peak $f_o$ decays through the transition regime (0.6 V < $V_{MG}$ < 1.6 V), replaced by $f_{o+i}$ as the dominant peak. We plot the corresponding area obtained from magnetic field frequency (Fig. 4e) and the plunger gate frequency (Fig. 4f), respectively, for each of the lowest-order peaks $f_o$, $f_i$, $f_{o+i}$, and $f_{o-i}$ as a function of $V_{MG}$. At the beginning of the transition regime where the ECs are not interacting, both $f_o$ and $f_{o+i}$ approach the corresponding AB frequency $\Phi_0^{-1} = e/h$ through the designed area. As $V_{MG}$ increases, however, $f_o$ stays nearly unchanged, while $f_{o+i}$ increases to reach the doubled value $2\Phi_0^{-1}$. The experimental observation that the dominant peak in the frequency-doubled regime corresponds to $f_{o+i}$ precludes the possibility of $2e$ charge pairing within the outer EC alone.

Instead, our frequency-doubled regime arises from Coulomb interaction between the spin-split ECs combined with charge quantization on the inner EC (Methods). Electrons would naturally tend to enter the inner EC at frequency $f_i$, but, due to charge quantization, cannot enter continuously. Hence, as the magnetic flux increases continuously, the area enclosed by the inner EC must shrink to maintain fixed charge. During this shrinking process, electron charge is transferred continuously into the interior, leaving missing electron charge between the outer and inner ECs. In the strongly coupled EC limit, this missing charge attracts an equal charge onto the outer EC for screening. In the absence of this screening effect, charge is continuously added to the outer EC with frequency $f_o$ according to the increased AB phase. In the coupled ECs, the combination of the screening-induced charge and the natural AB effect results in the outer EC charging at a frequency $f_{o+i}$. Therefore, the interference phase follows $f_{o+i}$. In addition to this continuous charging effect, electrons can tunnel into the inner EC from the external reservoirs. As previously discussed, each electron addition repels some electron charge from the outer EC, causing the negative interference phase shifts that we observed. For larger values of $V_{MG}$, as the bulk density increases, the inner and outer EC move closer together, and the system approaches the strong coupling limit, where the phase jumps are close to $-2\pi$ and unobservable, reflecting a full electron charge screening. Moreover, as the inner and outer ECs asymptotically enclose the same area, set by the confining potential of the device, the frequency $f_{o+i}$ approaches $2\Phi_0^{-1}$.

Note: a concurrent work also observed apparent AB frequency tripling, corresponding to the sum of the three $\nu = 3$ edge channel frequencies[38]. The framework that we developed here can be expected to naturally explain this observation, since in devices utilizing the graphene crystal edge, the sharp confining potential can lead to multiple ECs developing within a few magnetic lengths of the edge.[8] The combination of reduced spatial separation and reduced screening by nearby graphite gates may account for the observation of apparent tripling, arising from the outer EC screening both internal localized ECs.

**Conclusion and outlook**

We have investigated phase jumps and AB frequency modulation in a highly tunable graphene QHE FP interferometer with coupled ECs. We identify that interference phase jumps are related to the single electron charging events in the inner EC, and the transition of the AB frequency can be connected to the corresponding screening effect of the outer EC. As $V_{MG}$ increases, the EC coupling becomes strong and the AB frequency doubles, indicating a near-perfect screening between the ECs. Thus, our experimental observation supports the proposal that AB frequency doubling can be explained without explicitly introducing electron pairing within the outer two ECs[33]. In other words, a half flux quantum introduced in the two strongly coupled ECs can bring a full charge from the external reservoir and a $2\pi$ evolution of the observed interferometer phase.

Our observations do not exclude the possibility of further correlation effects in the strongly coupled ECs; the tunably coupled ECs discovered here provides a system to test the emergence of electron correlations in 1D systems[39]. However, AB frequency multiplication, which we explained within a single particle picture, cannot substantiate the correlation effect. Further experiments probing the strongly coupled limit, such as shot noise[25,40,41], finite-bias dependence[15], energy relaxation[42], and high-frequency transport[43–45] will provide further insight. More generally, inter-edge screening could affect interferometry in fractional fillings containing multiple ECs[46–48], and our versatile device will aid in controlling anyons in the fractional QHE.

**References**

1. Giamarchi, T. *Quantum Physics in One Dimension*. (Clarendon Press, 2003).
2. Wen, X.-G. *Quantum Field Theory of Many-body Systems: From the Origin of Sound to an Origin of Light and Electrons*. (Oxford University Press, 2007).
3. Marguerite, A. *et al.* Imaging work and dissipation in the quantum Hall state in graphene. *Nature* **575**, 628–633 (2019).
4. Uri, A. *et al.* Nanoscale imaging of equilibrium quantum Hall edge currents and of the magnetic monopole response in graphene. *Nat. Phys.* **16**, 164–170 (2020).
5. Halperin, B. I. Quantized Hall conductance, current-carrying edge states, and the existence of extended states in a two-dimensional disordered potential. *Phys. Rev. B* **25**, 2185–2190 (1982).
6. Chklovskii, D. B., Shklovskii, B. I. & Glazman, L. I. Electrostatics of edge channels. *Phys. Rev. B* **46**, 4026–4034 (1992).
7. Kim, S. *et al.* Edge channels of broken-symmetry quantum Hall states in graphene visualized by atomic force microscopy. *Nat Commun* **12**, 2852 (2021).
8. Coissard, A. *et al.* Absence of edge reconstruction for quantum Hall edge channels in graphene devices. *Science Advances* **9**, eadf7220 (2023).
9. Ji, Y. *et al.* An electronic Mach–Zehnder interferometer. *Nature* **422**, 415–418 (2003).
10. Bocquillon, E. *et al.* Electron quantum optics in ballistic chiral conductors. *Annalen der Physik* **526**, 1–30 (2014).
11. van Wees, B. J. *et al.* Observation of zero-dimensional states in a one-dimensional electron interferometer. *Phys. Rev. Lett.* **62**, 2523–2526 (1989).
12. de C. Chamon, C., Freed, D. E., Kivelson, S. A., Sondhi, S. L. & Wen, X. G. Two point-contact interferometer for quantum Hall systems. *Phys. Rev. B* **55**, 2331–2343 (1997).
13. Halperin, B. I., Stern, A., Neder, I. & Rosenow, B. Theory of the Fabry-Pérot quantum Hall interferometer. *Phys. Rev. B* **83**, 155440 (2011).

14. Nakamura, J., Liang, S., Gardner, G. C. & Manfra, M. J. Direct observation of anyonic braiding statistics. *Nat. Phys.* **16**, 931–936 (2020).
15. Nakamura, J., Liang, S., Gardner, G. C. & Manfra, M. J. Impact of bulk-edge coupling on observation of anyonic braiding statistics in quantum Hall interferometers. *Nat Commun* **13**, 344 (2022).
16. Carrega, M., Chirolli, L., Heun, S. & Sorba, L. Anyons in quantum Hall interferometry. *Nat Rev Phys* **3**, 698–711 (2021).
17. Feldman, D. E. & Halperin, B. I. Fractional charge and fractional statistics in the quantum Hall effects. *Rep. Prog. Phys.* **84**, 076501 (2021).
18. Ronen, Y. *et al.* Aharonov–Bohm effect in graphene-based Fabry–Pérot quantum Hall interferometers. *Nat. Nanotechnol.* **16**, 563–569 (2021).
19. Déprez, C. *et al.* A tunable Fabry–Pérot quantum Hall interferometer in graphene. *Nat. Nanotechnol.* **16**, 555–562 (2021).
20. Fu, H. *et al.* Aharonov–Bohm Oscillations in Bilayer Graphene Quantum Hall Edge State Fabry–Pérot Interferometers. *Nano Lett.* **23**, 718–725 (2023).
21. Zhao, L. *et al.* Graphene-Based Quantum Hall Interferometer with Self-Aligned Side Gates. *Nano Lett.* **22**, 9645–9651 (2022).
22. Zhang, Y. *et al.* Distinct signatures for Coulomb blockade and Aharonov-Bohm interference in electronic Fabry-Perot interferometers. *Phys. Rev. B* **79**, 241304 (2009).
23. Ofek, N. *et al.* Role of interactions in an electronic Fabry–Perot interferometer operating in the quantum Hall effect regime. *Proceedings of the National Academy of Sciences* **107**, 5276–5281 (2010).
24. Sivan, I. *et al.* Observation of interaction-induced modulations of a quantum Hall liquid's area. *Nat Commun* **7**, 12184 (2016).
25. Choi, H. K. *et al.* Robust electron pairing in the integer quantum hall effect regime. *Nat Commun* **6**, 7435 (2015).
26. Sivan, I. *et al.* Interaction-induced interference in the integer quantum Hall effect. *Phys. Rev. B* **97**, 125405 (2018).
27. Biswas, S., Kundu, H. K., Umansky, V. & Heiblum, M. Electron Pairing of Interfering Interface-Based Edge Modes. *Phys. Rev. Lett.* **131**, 096302 (2023).
28. Demir, A. *et al.* Correlated Double-Electron Additions at the Edge of a Two-Dimensional Electronic System. *Phys. Rev. Lett.* **126**, 256802 (2021).
29. Keimer, B., Kivelson, S. A., Norman, M. R., Uchida, S. & Zaanen, J. From quantum matter to high-temperature superconductivity in copper oxides. *Nature* **518**, 179–186 (2015).
30. Willett, R. L. The quantum Hall effect at 5/2 filling factor. *Rep. Prog. Phys.* **76**, 076501 (2013).
31. Li, J. I. A. *et al.* Even-denominator fractional quantum Hall states in bilayer graphene. *Science* **358**, 648–652 (2017).
32. Huang, K. *et al.* Valley Isospin Controlled Fractional Quantum Hall States in Bilayer Graphene. *Phys. Rev. X* **12**, 031019 (2022).
33. Frigeri, G. A., Scherer, D. D. & Rosenow, B. Sub-periods and apparent pairing in integer quantum Hall interferometers. *EPL* **126**, 67007 (2019).
34. Büttiker, M. Quantized transmission of a saddle-point constriction. *Phys. Rev. B* **41**, 7906–7909 (1990).
35. Zimmermann, K. *et al.* Tunable transmission of quantum Hall edge channels with full degeneracy lifting in split-gated graphene devices. *Nat Commun* **8**, 14983 (2017).
36. Feldman, D. E. & Halperin, B. I. Robustness of quantum Hall interferometry. *Physical Review B* **105**, 165310 (2022).
37. Röösli, M. P. *et al.* Observation of quantum Hall interferometer phase jumps due to a change in the number of bulk quasiparticles. *Phys. Rev. B* **101**, 125302 (2020).
38. Yang, W. *et al.* Evidence for correlated electron pairs and triplets in quantum Hall interferometers. Preprint at http://arxiv.org/abs/2312.14767 (2023).

39. Shavit, G. & Oreg, Y. Electron pairing induced by repulsive interactions in tunable one-dimensional platforms. *Phys. Rev. Res.* **2**, 043283 (2020).
40. Frigeri, G. A. & Rosenow, B. Electron pairing in the quantum Hall regime due to neutralon exchange. *Phys. Rev. Research* **2**, 043396 (2020).
41. Inoue, H. *et al.* Charge Fractionalization in the Integer Quantum Hall Effect. *Phys. Rev. Lett.* **112**, 166801 (2014).
42. Altimiras, C. *et al.* Non-equilibrium edge-channel spectroscopy in the integer quantum Hall regime. *Nature Phys* **6**, 34–39 (2010).
43. Bocquillon, E. *et al.* Separation of neutral and charge modes in one-dimensional chiral edge channels. *Nat Commun* **4**, 1839 (2013).
44. Freulon, V. *et al.* Hong-Ou-Mandel experiment for temporal investigation of single-electron fractionalization. *Nat Commun* **6**, 6854 (2015).
45. Hashisaka, M., Hiyama, N., Akiho, T., Muraki, K. & Fujisawa, T. Waveform measurement of charge- and spin-density wavepackets in a chiral Tomonaga–Luttinger liquid. *Nature Phys* **13**, 559–562 (2017).
46. Willett, R. L. *et al.* Interference Measurements of Non-Abelian e/4 & Abelian e/2 Quasiparticle Braiding. *Phys. Rev. X* **13**, 011028 (2023).
47. Kundu, H. K., Biswas, S., Ofek, N., Umansky, V. & Heiblum, M. Anyonic interference and braiding phase in a Mach-Zehnder interferometer. *Nat. Phys.* **19**, 515–521 (2023).
48. Nakamura, J., Liang, S., Gardner, G. C. & Manfra, M. J. Fabry-Pérot Interferometry at the ν=2/5 Fractional Quantum Hall State. *Phys. Rev. X* **13**, 041012 (2023).

## Methods

**Sample preparation.** The monolayer graphene stacks with hBN and graphite encapsulation used in this study were fabricated using the same polycarbonate (PC) polymer dry transfer method described in detail in our previous work (Ref. 19). The graphite top-gates and bottom-gate, which encapsulate the graphene channel after stacking, are crucial to screen charge disorder from the graphene channel, stabilizing robust integer and fractional QHE states at low magnetic fields (Supplementary Information). The stack used for all data shown here had a top (bottom) hBN thickness of 49 (27) nm. After adhering the stack to a substrate and annealing in vacuum at 300°C, the device geometry was defined by reactive ion etching in an inductively coupled plasma etching chamber using a polymethyl methacrylate (PMMA) resist patterned with electron-beam lithography as the etch mask. This etching was in two steps: first a pure 30W $O_2$ etch of the top graphite, then a 30W process $O_2$/$CHF_3$ to etch through the entire stack. Next, edge contacts to the exposed graphene were made by a 30W $CHF_3$ etch on the exposed hBN/graphene/hBN contact regions and thermal evaporating 2/7/150 nm of Cr/Pd/Au at an angle with rotation. Then, air bridge contacts were made to the top-graphite in various locations using a bilayer PMMA process followed by a short 20-25s 30W $O_2$ plasma PMMA residue clean and thermal evaporation of 2/7/350 nm Cr/Pd/Au. To etch the ~100 nm lines in the top graphite, a thinner PMMA resist was used and again a reactive ion etch with gentle 30W $O_2$ plasma alone was done in ~1 minute steps. In between etches, the two-probe resistance between each bridge-contacted gate was checked until they were all separated. Finally, bridge contacts to the separated central hexagon gate and suspended bridges over the QPC regions were deposited.

**Measurements.** The 8 top graphite gates in the device were separately controlled to set filling factors in each region at perpendicular magnetic field $B$, since Landau level filling factor (also simply called 'filling') $\nu \equiv n_e/n_\phi$, where $n_\phi = eB/h$ and $n_e$ is the areal electron density. At the region in the middle of the top-gate split-gates, where the graphite is etched away for a separation of ~150 nm, the electrostatics are tuned to create a saddle-point potential at the QPC. See Supplementary Information for details of this tuning process. Once an approximate saddle-point is formed at the QPCs using the graphite top-gates and bottom-gate, the suspended metal bridges over the QPCs are tuned to precisely set transmissions $T_{QPC1}$ and $T_{QPC2}$. The neighboring top-gates screen out stray fields generated by the suspended bridges such that $V_{QPC1}$ and $V_{QPC2}$ are primarily coupled to the graphene at the saddle-point of the QPCs. We interpret non-integer values $0 < T_{QPC} < 1$ as a transmission probability for electrons in the outer EC, which is partially transmitted, while for $1 < T_{QPC} < 2$, $T_{QPC} - 1$ gives the transmission probability for the inner EC.

Experiments were performed in an Oxford wet dilution system with base temperature ~20 mK and estimated ~20-25 mK electron temperature. The 24 DC measurement lines of the fridge were carefully thermalized through Thermocoax cables and 3 Sapphire plates between room temperature and the mixing chamber. A series of lumped element Pi and RC filters at the mixing chamber reduced electronic noise and ensured low electron temperature. Unless otherwise noted, a constant 6T perpendicular magnetic field was applied. Measurements were taken using standard low-frequency lock-in amplifier techniques with a typical AC excitation current of 1 nA at 17.77 Hz applied to the sample and simultaneously measured AC voltage drops and drained current. Graphite and suspended bridge gates were controlled with a house-made, low-noise 16-bit D/A voltage source. Bias dependence (see Supplementary Information) was taken by voltage biasing instead and adding in a DC bias at the source. Simultaneously, the DC voltage drop $V_D$ was measured on the same probes measuring the AC conductance so that the accurate voltage drop across the FP cavity was known. All data collected and analysis programs have been made available.

**Estimation of the coupling strength.** Although we have not attempted a detailed calculation of the coupling constants important for our analysis, we can at least advance some qualitative arguments for the trend that emerges from our analysis. The edge of the sample consists of alternating compressible and incompressible stripes whose width is set by electrostatics[6]. ECs are located in compressible stripes. It may be expected that the outermost EC is located along an electron density contour where the local Landau-level filling factor is ~ 0.5, while the second EC is located along a contour with filling ~1.5. Due to residual disorder and electron-electron interactions, the Hall plateau at $\nu = 2$ will set in when the bulk filling is smaller than 2, though larger than 1.5. The density profile produced by charges on confining gates should be relatively smooth, so that the spatial separation between the outer most EC and the second EC should be relatively large at this point, and the Coulomb coupling between the channels, screened by the gates, should be relatively weak. As the electron density is increased, the inner EC should move closer to the outer edge, and the coupling should become stronger, and it is plausible that by the time the device enters the $\nu = 3$ plateau, the value of $K_{12}/K_1$ is close to 1.

Further increases in the density should produce additional ECs, which are totally reflected at the QPCs and do not contribute directly to the transport. The number of electrons on any additional closed ECs, as on other localized states, will be restricted to integer values, and in principle, due to Coulomb interactions, there should be a jump in the interference phase of outer edge states each time this integer changes by one. However, Coulomb interactions in our system are strongly screened by the nearby gates, so if the additional channels are not close to the outer two ECs, the jumps would be too small to be observable. In monolayer graphene, the energy gap at $\nu = 2$, which is due to the cyclotron energy, is much larger than the gaps at $\nu = 1, 3, 4$, and 5, which arise from electron-electron interactions. Consequently, we expect that the spatial separation between the outermost EC and the second EC will tend to be small compared to the separation between the second EC and any additional ECs.

Another issue is the stability criterion embodied in the requirement $|K_{12}|^2 \leq K_1 K_2$. This requirement is automatically satisfied if we assume that when the two outer ECs are close together, the energy for adding an electron to either one of them is dominated by an electrostatic energy that depends primarily on the total charge on the edges, and only weakly on the difference between them, so that $E = a\,\delta Q_1^2 + b\,\delta Q_2^2 + 2K_{12}(\delta Q_1 + \delta Q_2)^2$, with $a$ and $b$ small compared to $K_{12}$. Then, $K_1$ and $K_2$ will be approximately equal to each other and slightly larger than $K_{12}$.

This analysis is compatible with experiments in GaAs interferometers where the ECs occur at the boundary between two QHE states of different integer filling fractions (Ref. 27). There it was found that the $h/2e$ periodicity occurred only if the outer EC and second EC belong to the same orbital Landau level, and not if they belong to different levels. In the first case, the energy gap for the QHE state between the two ECs will arise from electron-electron interactions, while the energy gap in the second case will be dominated by the generally larger cyclotron energy. Therefore, in the first case, when the density is increased enough to populate a third QHE state in the bulk of the sample, the two outer ECs might be pushed so close to each other that they are strongly coupled, while this might not be expected to happen in the second case.

**Physics of AB frequency doubling at strong coupling.** The meaning of the charge fluctuations $\delta Q_1$ and $\delta Q_2$ can be made more precise as follows. As stated in the main text, we define $Q_1$ as the number of electrons in the lowest spin-split Landau level enclosed by the outer edge mode and $Q_2$ as the number of electrons in the higher spin state enclosed by the inner mode. These charges are related to the enclosed areas $A_1$ and $A_2$ by $Q_i = A_i B/\Phi_0$, where $i = 1$ or 2. These areas are allowed to deviate slightly from the ideal areas $\bar{A}_i$, which are assumed to be smooth functions of $V_{\text{PG}}$ and, at most weakly varying functions of $B$ and $V_{\text{MG}}$. Then $\delta Q_i = Q_i - B\bar{A}_i/\Phi_0$, and the energy $E$ may be expanded to quadratic order in $\delta Q_i$ as stated above.

When the inner mode is completely reflected at the QPC, the charge $Q_2$ is constrained to be an integer, while the charge $Q_1$ can change continuously, assuming that the outer edge is mostly transmitted through the QPCs. At low temperatures the charges will be determined so as to minimize $E$, subject to the integer constraint.

If $Q_2$ is held fixed while the magnetic field is increased by a small amount $dB$, the inner edge charge $\delta Q_2$ will change by an amount $-dB\bar{A}_2/\Phi_0$. This happens because, as the area shrinks, charge is transferred from the edge region to the interior, where it is effectively screened by the gates, leaving a charge deficit at the edge. In the strong coupling limit, this will cause $\delta Q_1$ to increase by an equal amount. Thus, the total charge $Q_1$ in the lowest spin-split Landau level will increase by $dQ_1 = dB(\bar{A}_1 + \bar{A}_2)/\Phi_0$, and the interferometer phase $\theta$ will increase by $2\pi dQ_1$.

If $B$ is increased by a large amount, the value of $Q_2$ will not be fixed but will undergo periodic integer jumps. In the strong coupling limit, the jump in $Q_1$ caused by a jump in $Q_2$ will also be an integer. This will cause $\theta$ to jump by a multiple of $2\pi$, which will be invisible in an interferometer experiment. Thus, the observed rate of change of the phase will be $d\theta/dB = 2\pi(\bar{A}_1 + \bar{A}_2)/\Phi_0$, which is equal to $4\pi\bar{A}_1/\Phi_0$, if we neglect the difference between $\bar{A}_1$ and $\bar{A}_2$. This rate of change is twice as fast as would have been observed in the absence of coupling between the inner and outer edge modes.

We remark that in the course of adding one flux quantum to the area $\bar{A}_1$, one would expect on average to have a jump by one electron in each spin state. So, in general, one will have one positive jump in $Q_2$ and one negative jump in $Q_1$. Thus, while the observed interference phase will change by an amount equivalent to a change of two electrons, the actual change in $Q_1$ will only be one electron.

**Robustness of the theoretical predictions.** As discussed in Ref. 36, when a single EC passes through the two constrictions, with weak backscattering at the constrictions, the interference phase seen at low temperatures and low source-drain voltage is given by $\theta = 2\pi Q + \theta_0$, mod $2\pi$, where $Q$ is the total electron charge (in units $e$) in the region between the two scattering points (the expectation value of the charge on the interferometer in its ground state) and $\theta_0$ is a constant for small variations in $B$, $V_{PG}$, and $V_{MG}$. The argument is essentially the same if the backscattering is not weak. The principal effect of stronger backscattering at the QPCs is to add a term to the energy $E$ that favors integer values of the charge $Q_1$ and hence integer values of the total charge on the interferometer. This means that as the control parameters are varied continuously, the phase difference $\theta - \theta_0$ will undergo an additional modulation pulling it towards the nearest integer multiple of $2\pi$. If we define $\theta_b$ as the value of the interferometer phase that would occur in the limit of weak backscattering, for the given value of the control parameters, then the actual value of $\theta$ should have the form $\theta = \theta_b + \delta\theta$, where $\delta\theta$ is a periodic function of $\theta_b - \theta_0$. In addition, in the presence of finite back scattering, interference contribution to the measured resistivity may no longer be a simple sinusoidal function of $\theta$ but can contain higher harmonics. The combination of these effects means that the interference current will remain a periodic function of $\theta_b$, with period $2\pi$, but the relative amplitudes of various harmonics may be modified. In the main text, it was argued that $\cos\theta_b$ should be a two-dimensional periodic function of $B$ and the gate voltages, with frequencies expressed in terms of two non-colinear fundamental vectors in reciprocal parameter space. The effect of finite backscattering at the QPCs will be to modify the amplitudes of the various Fourier components, but not to change their positions.

Using similar arguments, we may argue that measurement at finite temperature should not change the locations of the fundamental frequency vectors, but thermal fluctuations will reduce the Fourier amplitudes. In general, at high temperatures $T$, the amplitude of a given Fourier component will fall off, proportional to $e^{-T/\varepsilon}$, where $\varepsilon$ will be different for each Fourier component. At sufficiently high temperatures, therefore, only the one or two components with the largest values of $\varepsilon$ will remain visible. The values of $\varepsilon$ will depend on details of the system, but typically the Fourier components that are most prominent at $T = 0$ will be the ones that persist to highest temperatures.

For our system, in the case where there is only a single EC, as we find for bulk filling less than 2, the value of $\varepsilon$ for the lowest Fourier mode is predicted to be $\varepsilon = hv/(2\pi^2 P)$, where $v$ is the EC velocity and $P$ is the perimeter of the interferometer path. For the case of two strongly coupled edge channels, the prediction is $\varepsilon = hv/(4\pi^2 P)$, where $v$ is now the velocity of the fast charge mode. In both cases, the dominant effects come from thermal fluctuations $e\delta Q$ of the charge on the edge, whose energy cost is given by $(e\delta Q)^2/(2\gamma P)$, where $\gamma$ is the capacitance per unit length of the edge. The velocity $v$ is given by $v = \delta\sigma_{xy}/\gamma$, where $\delta\sigma_{xy}$ is the change in Hall conductance across the edge. Using our lithographically defined perimeter $P = 4.24$ μm and the velocity $v_e = \frac{e\Delta V_D P}{h} = 1.46 \times 10^5$ m/s extracted from finite-bias dependence in the uncoupled case (SI), we find $\varepsilon = 83.7$ mK, well above our estimated electron temperature.

**Acknowledgements**

We thank Andrew Pierce and Raymond Ashoori for helpful comments in the early stages of this work. We also thank Raymond Ashoori for lending important cryostat parts used in this measurement and Jim MacArthur for building electronics used in our experiment. P.K., T.W., and Y.R. acknowledge support from DOE (DE-SC0012260) for sample preparation, measurement, characterization, and analysis. J.R.E. acknowledges support from ARO MURI (N00014-21-1-2537) for sample preparation, measurement, characterization, and analysis. K.W. and T.T. acknowledge support from the Elemental Strategy Initiative conducted by the MEXT, Japan, Grant Number JPMXP0112101001, JSPS KAKENHI Grant Number JP20H00354 and the CREST(JPMJCR15F3), JST. D.E.F. and Z.W. acknowledge support by the National Science Foundation under Grant No. DMR-2204635. B.I.H. acknowledges support from NSF grant DMR-1231319. Nanofabrication was performed at the Center for Nanoscale Systems at Harvard, supported in part by an NSF NNIN award ECS-00335765.

**Author contributions**

T.W. and D.N. stacked the graphite-encapsulated heterostructures. T.W. performed the nanofabrication, measurements, and data analysis. J.R.E. and Y.R. assisted in the measurement and analysis. D.E.F., B.I.H., and Z.W. contributed the theoretical analysis. M.E.W. and A.Y. provided the measurement cryostat and collaborated on the discussions and analysis. K.W. and T.T. provided the hBN crystals. T.W., B.I.H., J.R.E., and P.K. wrote the paper with input from all authors.

**Competing interests**

The authors declare no competing interests.

# Figures

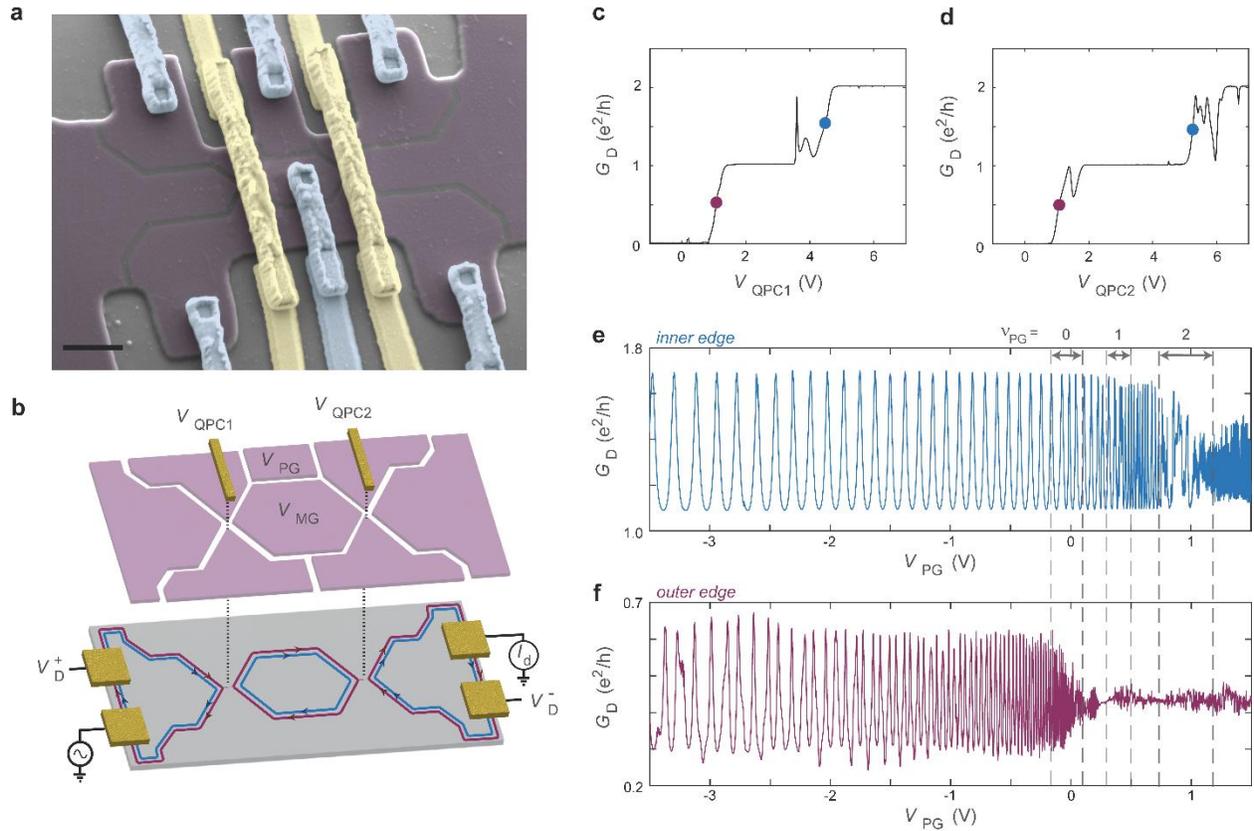

**Fig. 1 | Highly tunable Fabry-Pérot interferometer in graphene. a,** False-colour scanning electron microscopy image of a FP device identical to the device measured here. The graphite top-gate layer is selectively etched to form 8 separated top-gates (purple). Metal bridges (blue) connect to each graphite top-gate region and two additional bridges (yellow) suspend over the QPCs. The lithographic area of the interferometer cavity (area $A = 1.16$ μm$^2$) is defined by the central hexagonal top-gate. Scale bar: 1 μm. **b,** Schematic diagram of a FP at $\nu = 2$ illustrating interference of the outer EC (red) while the inner EC (blue) forms a closed annulus inside the FP. Voltages applied to the suspended metal bridges $V_{\text{QPC1}}$ and $V_{\text{QPC2}}$ selectively gate the QPC constrictions through the etched graphite gaps. We measure the diagonal conductance $G_{\text{D}} = (V_{\text{D}}^+ - V_{\text{D}}^-)/I_{\text{d}}$, where $V_{\text{D}}^\pm$ and $I_{\text{d}}$ are measured voltages in ($\pm$) probes and drained current, respectively. See SI for more device details. In addition to magnetic field, we tune the interference phase using voltage $V_{\text{MG}}$ on the 'middle gate' or $V_{\text{PG}}$ on the 'plunger gate'. **c,** Conductance as a function of $V_{\text{QPC1}}$ with $V_{\text{QPC2}} = 7$ V (i.e. open with $T_{\text{QPC2}} = 2$) demonstrating QPC1 tunings to interfere outer EC (red dot) and inner EC (blue dot) in $\nu = 2$. **d,** Same type of plot as **c**, but demonstrating QPC2 operation instead of QPC1. See SI for QPC tuning details and voltages set on the other gates to form QPC saddle-points. **e-f,** Characteristic FP oscillations as a function of $V_{\text{PG}}$ for the inner EC and outer EC, respectively, at the indicated QPC tunings. All data is at fixed magnetic field $B = 6$ T.

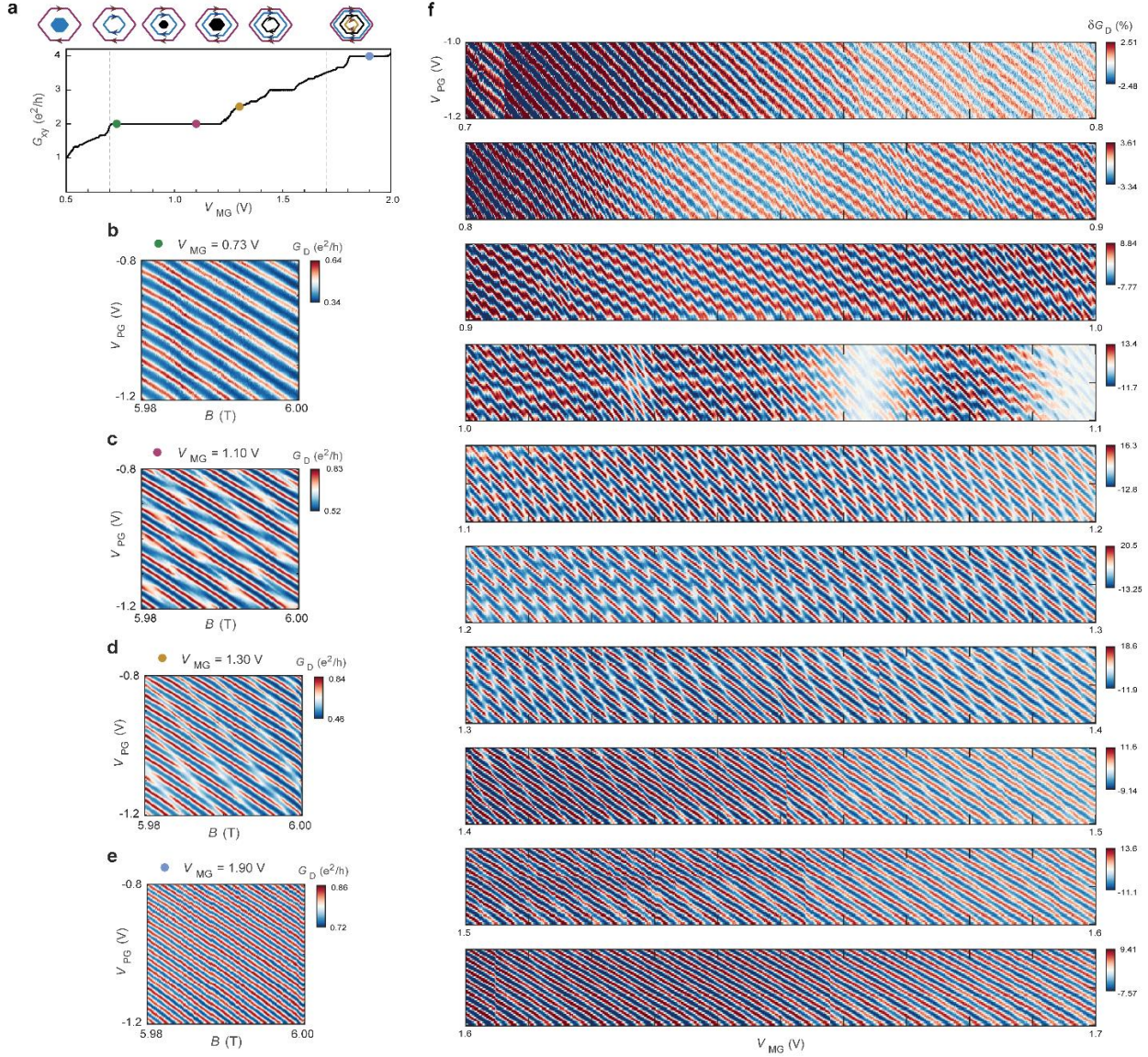

**Fig. 2 | AB oscillation frequency doubling transition of outer EC tuned with $V_{MG}$. a,** Hall conductance $G_{xy}$ of the device with both QPCs tuned to be fully open, demonstrating that $V_{MG}$ tunes the filling $\nu$ of the FP at a fixed magnetic field $B = 6$ T. Colored dots indicate points at which interference data are shown in **b-e** while vertical dashed lines show the range of $V_{MG}$ swept for **f**. Top inset pictures illustrate the corresponding compressible regions expected in the FP cavity. **b-e,** Conductance $G_D$ oscillations on the outer EC with $V_{PG}$ and $B$, for each of the indicated $V_{MG}$ values. **f,** Conductance $G_D$ oscillations on the outer EC with $V_{PG}$ and $V_{MG}$, for $V_{MG}$ swept continuously over the transition from apparent $h/e$ to $h/2e$ oscillations periodicity, at $B = 6$T. Here we plot $G_D$ as a percentage of $\frac{e^2}{h}$ deviation from the average value, which is calculated for each fixed $V_{MG}$ linecut and subtracted off. QPCs are retuned to maintain $T_{QPC1} = T_{QPC2} = 0.5$ over the dataset. We do not observe further phase jumps or periodicity changes past $V_{MG} \approx 1.7$ V (checked up to up $V_{MG} = 3.2$ V, corresponding to $\nu = 7$).

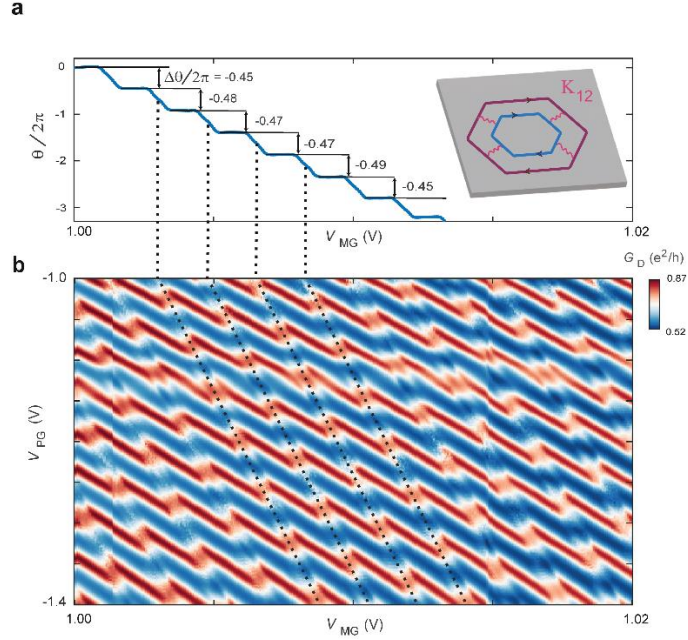

**Fig. 3 | Phase jump extraction in the transition regime. a,** Phase of the 1D fast Fourier transform (FFT) extracted along linecuts parallel to the phase jumps in **b**. The phase is evaluated at the dominant frequency in the FFT amplitude spectrum for the linecuts in between phase jumps. A linear increase in phase extracted from the regions without phase jumps is subtracted off to make the phase jump magnitude evident as the vertical shift between plateaus in this plot. From this data we extract $\Delta\theta/2\pi \approx -0.47$, reflecting approximately half of an electron repelled from the outer EC for each charge added to $Q_2$. Inset: illustration of the coupling $K_{12}$ between the outer and inner ECs contributing to the phase jumps. **b,** Conductance $G_D$ oscillations on the outer EC with $V_{PG}$ and $V_{MG}$ near the center of the transition regime showing periodic phase jumps along the dashed black lines. Note that increasing $V_{MG}$ adds electrons to the system or equivalently increases phase, so the phase jumps correspond to negative shifts in phase i.e., repulsion of electrons from the FP cavity.

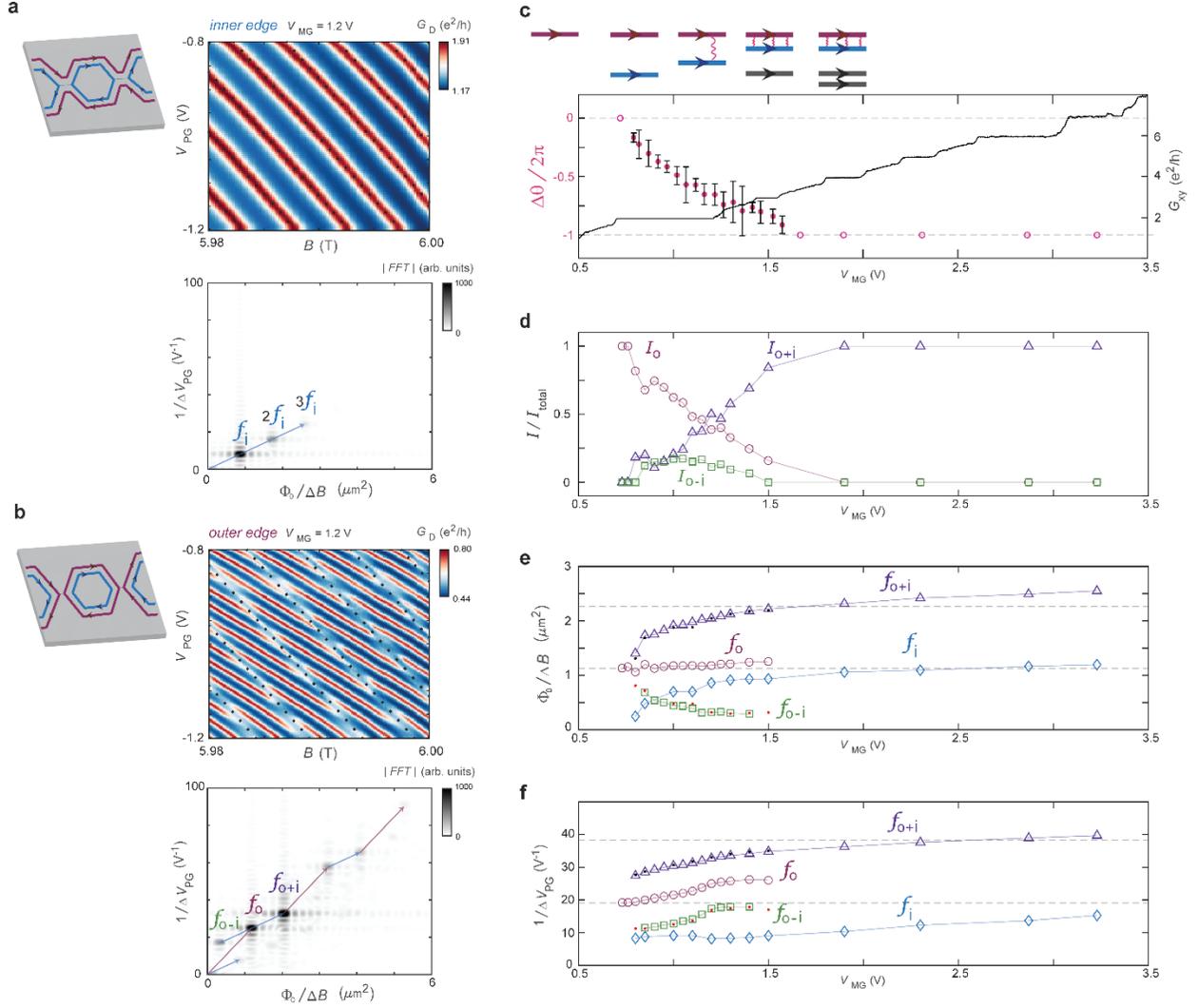

**Fig. 4 | Comparison of inner and outer EC interference and couplings across transition. a,** Conductance $G_D$ oscillations on the inner EC ($T_{QPC1} = T_{QPC2} = 1.5$) with $V_{PG}$ and $B$, for $V_{MG} = 1.2$V. Left: illustration of interference on inner EC. Bottom: 2D FFT of the $G_D$ oscillations showing peak $f_i$ and its harmonics. **b,** Same analysis and $V_{MG}$ value as in **a** but for interference on the outer EC ($T_{QPC1} = T_{QPC2} = 0.5$), showing the peaks $f_o$, $f_{o+i}$, and $f_{o-i}$ and their harmonics. **c,** Magnitude of the phase jump (obtained using the method shown in Fig. 3) as a function of $V_{MG}$, showing that it is continuously tunable. Each data point is averaged over $\sim$0.25 V range in $V_{MG}$ and error bars indicate $\pm 1$ standard deviation over the phase jumps detected in this range. We show $G_{xy}$ of the device, reflecting $\nu$, for reference. Open circle data points represent zero observable phase jumps over the corresponding $V_{MG}$ range, hence we infer a magnitude of 0 or $-1$. **d,** Normalized magnitudes $I_o$, $I_{o+i}$, and $I_{o-i}$ of the respective peaks $f_o$, $f_{o+i}$, and $f_{o-i}$ obtained as a function of $V_{MG}$. $I_o$, $I_{o+i}$, and $I_{o-i}$ are normalized by the sum $I_o + I_{o+i} + I_{o-i}$ to show their relative contributions. We extract each data point from a 2D dataset like panel **b**, a subset of which are shown in Extended Data Fig. 2. **e,** Magnetic field frequency converted to area for peaks $f_o$, $f_i$, $f_{o+i}$, and $f_{o-i}$ tracked through the transition. Note that $f_i$ is measured from a separate measurement of interference on the inner EC (Extended Data Fig. 3), while the other peaks are all extracted from interference on the outer EC. **f,** Same as **e** but for plunger gate frequency. Horizontal dashed lines in **e-f** indicate the corresponding $f_o$ and $2f_o$ values before the transition. Black (red) dots show calculated $f_o \pm f_i$ from outer and inner EC data, which match the peaks identified as $f_{o+i}$ and $f_{o-i}$, respectively.

**Extended data figures**

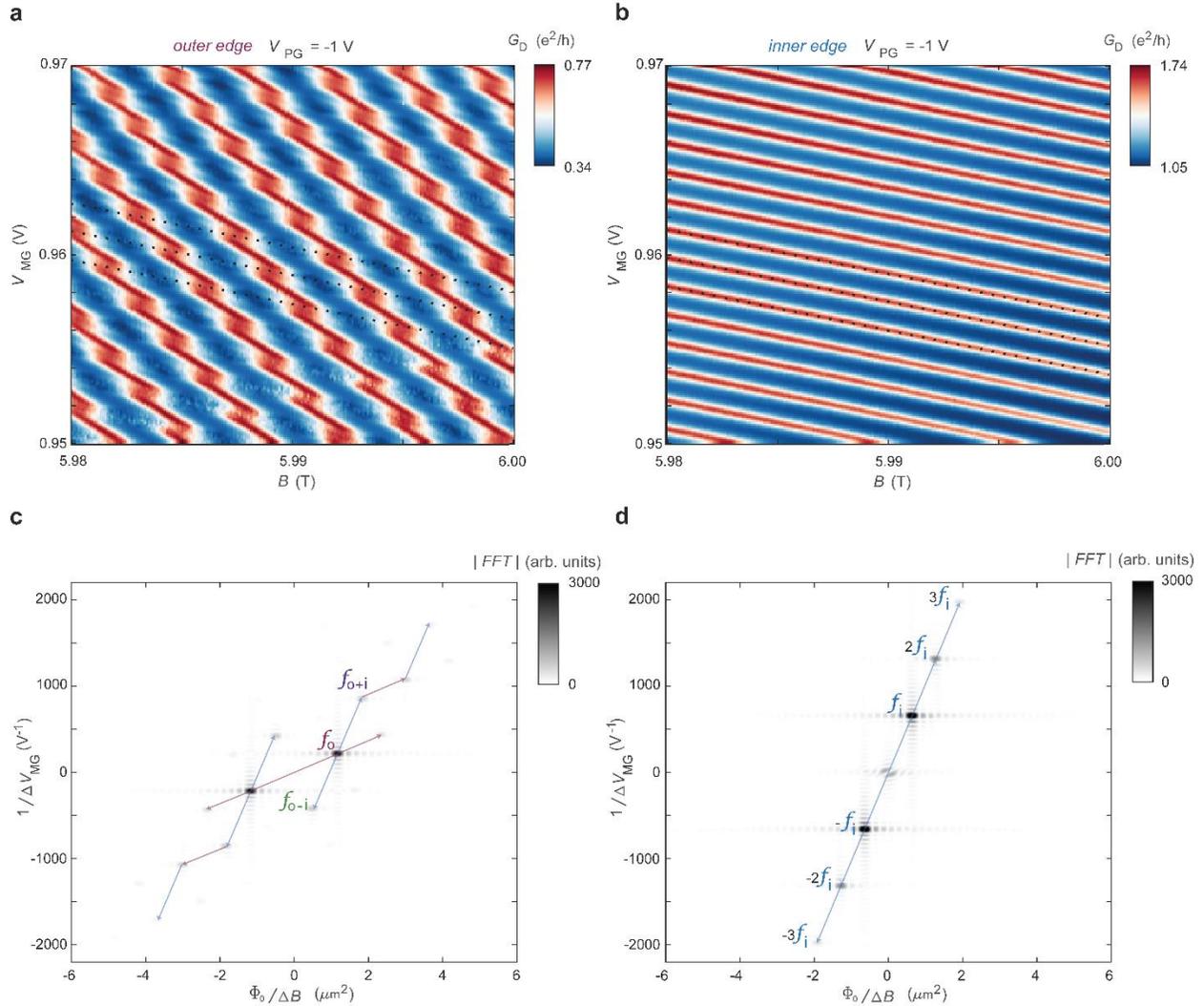

**Extended Data Fig. 1 | Comparison of the outer and inner EC interference for different parameters. a,** Conductance $G_D$ oscillations on the outer EC with $V_{MG}$ and $B$, for the indicated $V_{PG}$ value. Dashed black lines indicate the visible phase jumps in the data. **b,** Conductance $G_D$ oscillations on the inner EC with $V_{MG}$ and $B$. Dashed black lines indicate the oscillation maxima, which match the same slope as in **a**. **c,** 2D FFT of the $G_D$ oscillations from panel **a** showing the peaks $f_o$, $f_{o+i}$, $f_{o-i}$, and their harmonics. **d,** 2D FFT of the $G_D$ oscillations from panel **b** showing the peaks $f_i$ and its harmonics. The lattice of peaks in **c** consists of sums and differences of the fundamental frequencies of the outer EC $f_o$ and the inner EC $f_i$. Note that though the relative couplings to the parameters shown here are different than in Fig. 4, the same logic applies to understanding the peaks.

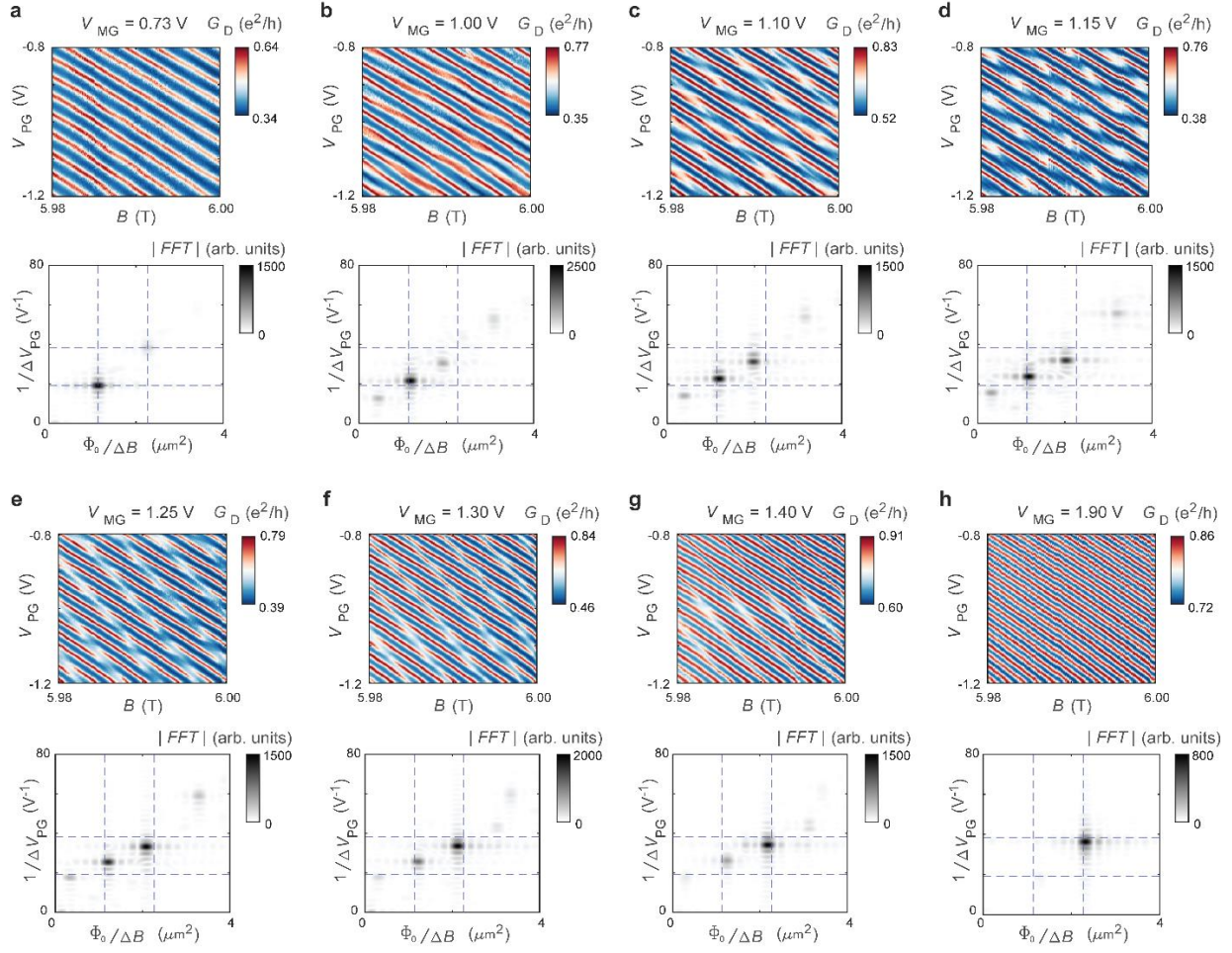

**Extended Data Fig. 2 | Magnetic field dependence of outer EC interference through the periodicity transition. a-h,** Top: conductance $G_D$ oscillations on the outer EC with $V_{PG}$ and $B$, for the indicated $V_{MG}$ values through the periodicity transition. Bottom: 2D FFT of the $G_D$ oscillations from the top panels showing the peaks $f_o$, $f_{o+i}$, $f_{o-i}$ continuously evolving through the transition. Horizontal dashed lines indicate $1/\Delta V_{PG} = 19.2\ \text{V}^{-1}$ and $2 \times 19.2\ \text{V}^{-1}$, corresponding to the frequencies $f_o$ and $2f_o$ before the transition. Similarly, vertical dashed lines indicate $\Phi_0/\Delta B = 1.13\ \mu\text{m}^2$ and $2 \times 1.13\ \mu\text{m}^2$. The designed FP area $A = 1.16\ \mu\text{m}^2$.

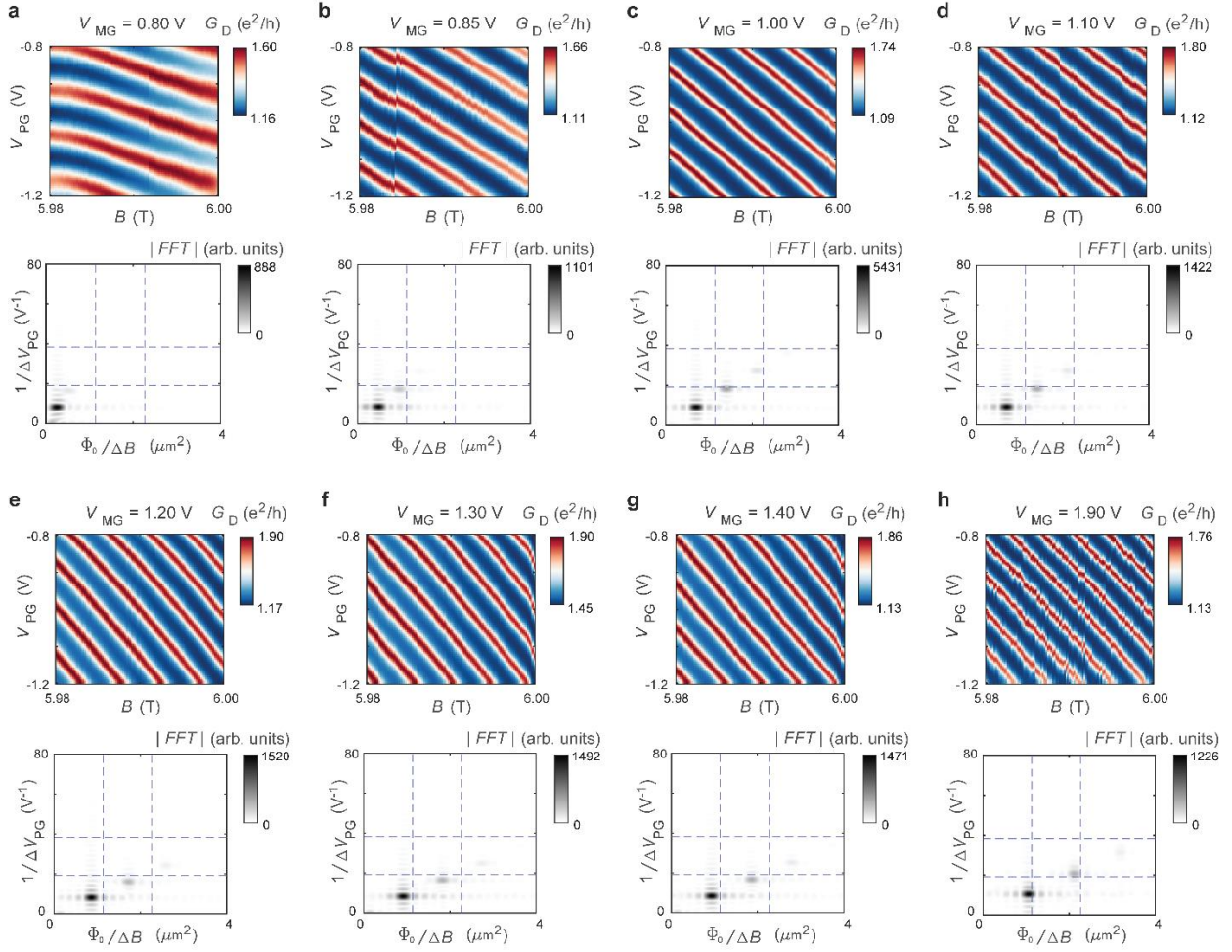

**Extended Data Fig. 3 | Magnetic field dependence of inner EC interference through the periodicity transition.** **a-h,** Top: conductance $G_D$ oscillations on the inner EC with $V_{PG}$ and $B$, for the indicated $V_{MG}$ values through the periodicity transition. Bottom: 2D FFT of the $G_D$ oscillations from the top panels showing the peaks $f_i$ and its harmonics continuously evolving through same range where the interference on the outer EC undergoes the transition. Horizontal dashed lines indicate $1/\Delta V_{PG} = 19.2\ \text{V}^{-1}$ and $2 \times 19.2\ \text{V}^{-1}$, corresponding to the frequencies $f_o$ and $2f_o$ before the transition. Similarly, vertical dashed lines indicate $\Phi_0/\Delta B = 1.13\ \mu\text{m}^2$ and $2 \times 1.13\ \mu\text{m}^2$. The designed FP area $A = 1.16\ \mu\text{m}^2$. The area of the inner EC evidently starts from near zero and increases monotonically, approaching the same area of the outer EC, which is roughly bound by the lithographic area. Note also that phase jumps are not observed on the inner edge.

# Supplementary Information

## DEVICE CHARACTERIZATION AND QPC OPERATION

Fig. S1a/b shows the general Hall characterization of the device, measured on the across the FP cavity as a typical Hall bar, with all top gates grounded. Fig. S1c shows two linecuts at B = 6 with all top gates grounded (0V) or set to 1V, showing the line of constant filling for the center of the plateaus. This provides the required compensating top-gate voltages that must be swept to keep $\nu = 2$ in the relevant interferometer regions (see Fig. 1a/b) when the various top gates are tuned to change the QPC transmissions (Fig. S2).

Fig. S2a/b shows the individual left/right QPC transmission given that the right/left QPC is fully open. The (left, middle)/(right, middle) top gates are swept along with the back gate on the x-axis to keep $\nu = 2$ to the sides of the QPC. The y-axis shows the sweep of the split gate voltages. Together, these two axes effectively tune the QPC and split gate filling factors related by a linear transformation. The new axes that correspond to these more physically relevant variables are indicated by the dashed lines which separate regions of different integer filling, with the x'-axis corresponding to the QPC filling (green lines) and the y'-axis that of the split gates (grey lines).

Critical to the measurements in the main text, we have added metallic QPC bridge gates to our previous device design[1] (Fig. 1a/b). This allows for the precise tuning of the exposed QPC regions in the etched top gate trenches without tying the back or split gates to specific values. This gives the device an increased degree of freedom for additional tunability across a broader range of gate voltages. For all data shown in the main text, the back gate and split top-gates are set to the values indicated by the blue dots in each plot, and the QPC filling (transmission) is then tuned separately by sweeping the bridge gate voltages. Importantly this corresponds to a regime where the split gate fillings are $\nu = 0$ to disallow any additional modes transmitting or tunneling through the barriers.

## WEAK AND STRONG BACKSCATTERING LIMITS

Fig. S3 shows diagonal interference data for weak (a) and strong (b) backscattering limits. Interferometer devices are typically measured in the weak backscattering limit where the interference signal can be interpreted as single-particle interference from a sum of particle trajectories transmitting/reflecting through the cavity. Fig. S3a shows the AB interference signal within the intermediate inter-edge coupling regime showing the phase jumps discussed in the main text ($T_{\text{QPC1}} = T_{\text{QPC2}} = 0.9$). However, when the QPCs are pinched off ($T_{\text{QPC1}} = T_{\text{QPC2}} = 0.01$), such as in Fig. S3b, the center cavity becomes analogous to a quantum dot where charge tunneling dominates transmission. While the signal is heavily suppressed, still visible is the same phase-jump periodic pattern of the weak backscattering AB data (Fig. S3a). In this strong backscattering regime, a capacitively-coupled double quantum dot model[2] may be used to explain the charge configurations localized on the inner and outer annular EC compressible regions, which due to the charge-phase relation discussed in the main text is related to the interference pattern observed at weak-backscattering.

## SINGLE-EDGE INTERFERENCE WITH DUAL-EDGE CAVITY

Fig. S4 shows data similar to Extended Data Fig. 2, except with the left/right regions astride the FP cavity tuned to $\nu = 1$. The center cavity remains in $\nu = 2$, thus the single incident edge is partially transmitted into the outer of the two edges within the cavity. Fig. S4b-d shows the evolution of phase jumps corresponding to inter-edge coupling discussed in the main text. The corresponding 2D FFTs show qualitatively identical behavior to Extended

Data Fig. 2 with the $f_o$, $f_{o+i}$, $f_{o-i}$ peaks evolving into the final approximate $2f_o$ signal. The observations in the main text are thus unaffected by the external filling factor and intrinsic to the interferometer region.

TEMPERATURE AND FINITE SOURCE-DRAIN BIAS DEPENDENCE

Fig. S5 compares the diagonal conductance oscillations in the main text's typical operating regime ($\nu = 2$, interfering the outer EC) at two different temperatures. The phase jumps corresponding to inter-edge coupling maintain the same qualitative behavior with an identical periodicity between 60mK and 440mK. The visibility is drastically reduced by 440mK, yet phase jump magnitude remains relatively unchanged, reflecting the fact that the inter-edge Coulomb coupling and ground-state charge configurations are zero temperature properties.

Lastly, we measure finite-bias dependence of the outer EC interference. We observe a typical bias dependence in the uncoupled case (Fig. S6a), consistent with a highly asymmetrical voltage drop across the FP cavity[3]. The bias voltage spacing of visibility nodes $\Delta V_D = 142$ μV and designed perimeter of the FP cavity $P = 4.24$ μm suggest an edge mode velocity $v_e = e\Delta V_D P/h = 1.46 \times 10^5$ m/s, consistent with previous observations in graphene[1,3] and GaAs[4,5] based FP interferometers. The intermediate coupling regimes Fig. S6b-c show complicated behavior which we cannot understand without a detailed theoretical model. However, in the strongly coupled limit, Fig. S6d, we again observe a relatively simple pattern corresponding to the $f_{o+i}$ frequency. The outer node spacing is similar to Fig. S6a, but there is a reduced-width central node, a signature which may indicate chiral Luttinger liquid physics[4]. Interpreting such data requires further experimental and theoretical work.

**References**

1. Ronen, Y. *et al.* Aharonov–Bohm effect in graphene-based Fabry–Pérot quantum Hall interferometers. *Nat. Nanotechnol.* **16**, 563–569 (2021).
2. Kouwenhoven, L. P. *et al.* Electron Transport in Quantum Dots. in *Mesoscopic Electron Transport* (eds. Sohn, L. L., Kouwenhoven, L. P. & Schön, G.) 105–214 (Springer Netherlands, 1997). doi:10.1007/978-94-015-8839-3_4.
3. Déprez, C. *et al.* A tunable Fabry–Pérot quantum Hall interferometer in graphene. *Nat. Nanotechnol.* **16**, 555–562 (2021).
4. Nakamura, J., Liang, S., Gardner, G. C. & Manfra, M. J. Impact of bulk-edge coupling on observation of anyonic braiding statistics in quantum Hall interferometers. *Nat. Commun.* **13**, 344 (2022).
5. Nakamura, J. *et al.* Aharonov–Bohm interference of fractional quantum Hall edge modes. *Nat. Phys.* **15**, 563–569 (2019).

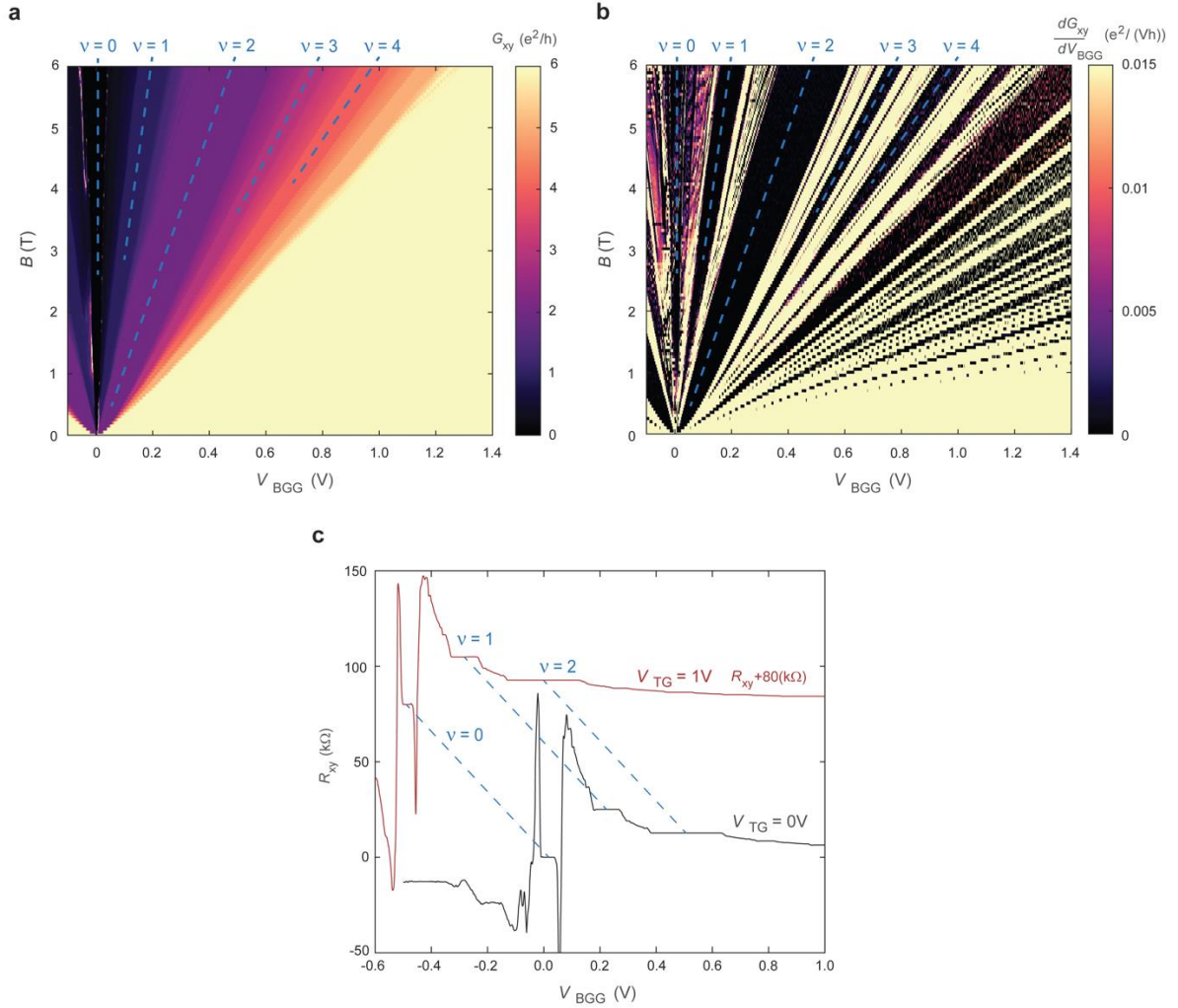

**FIG. S1. Device quantum Hall characterization. (a)** Hall conductance of [RHS/LHS/config] of device in operation regime of main text with density tuned via the graphite back gate. **(b)** Derivative of Hall conductance in (a) to emphasize the flat plateaus, including well-developed broken symmetry integer and fractional QHE states. **(c)** Two transverse Hall resistance measurements at different top gate voltages at $B = 6$T. Lines corresponding to the center of plateaus superimposed, indicating the required back gate and top gate voltages extrapolated to keep constant filling.

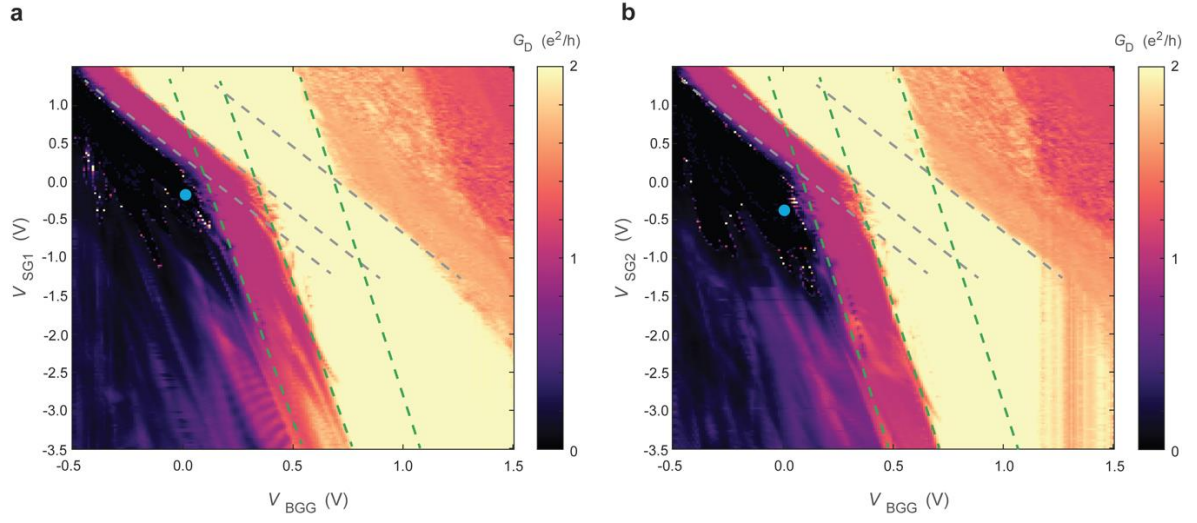

**Fig. S2. QPC tuning via back and split gates.** Individual transmissions of the left **(a)** and right **(b)** QPCs. For the graphite back gate sweep on the x-axis, the (left, middle)/(right, middle) graphite top gates are swept to compensate to keep the filling $\nu = 2$ underneath according to constant filling lines fit from scans like Fig. S1c. The dashed lines indicate steps in integer filling along the transformed axes corresponding to the QPC regions (green lines) and split gate regions (grey lines). Blue dots on each plot indicate the operational point of the interferometer for all interference data shown in the main text, effectively setting the split gate filling to 0 while the QPC bridge gates are swept to vary the QPC transmissions.

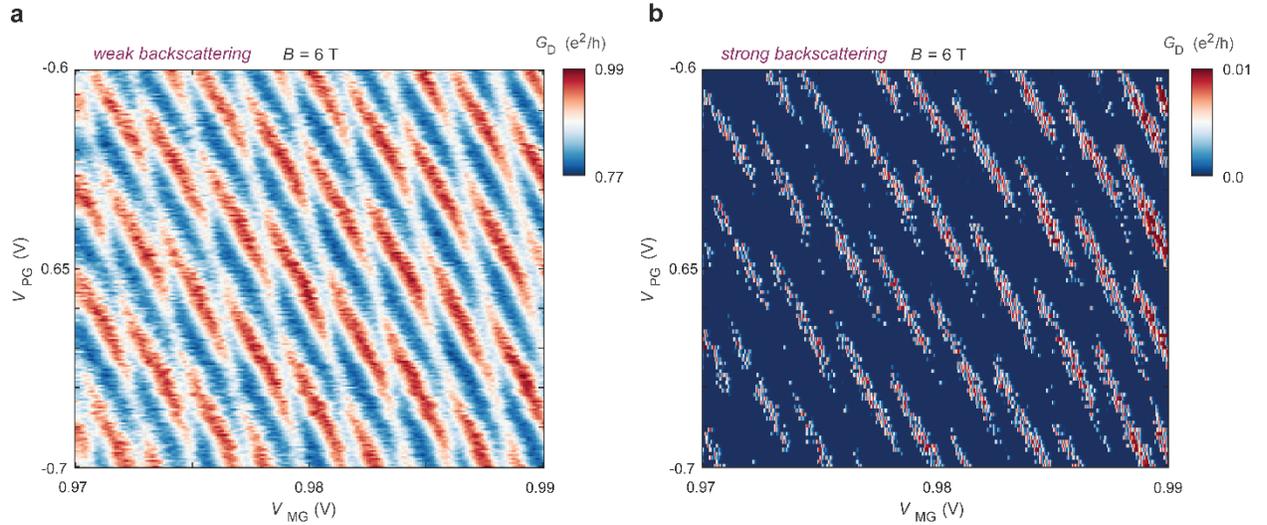

**Fig. S3. Weak and strong backscattering limits of intermediate-coupling regime.** Diagonal conductance of outer EC through the FP cavity with the QPCs tuned to the weak ($T_{QPC1} = T_{QPC2} = 0.9$) backscattering **(a)** and strong ($T_{QPC1} = T_{QPC2} = 0.01$) backscattering **(b)** limits. The typical AB interference signal as shown in the main text is seen in (a) along with phase jumps corresponding to inter-edge coupling, while a similar periodic pattern is seen in (b) where each QPC is functioning as a tunnel barrier.

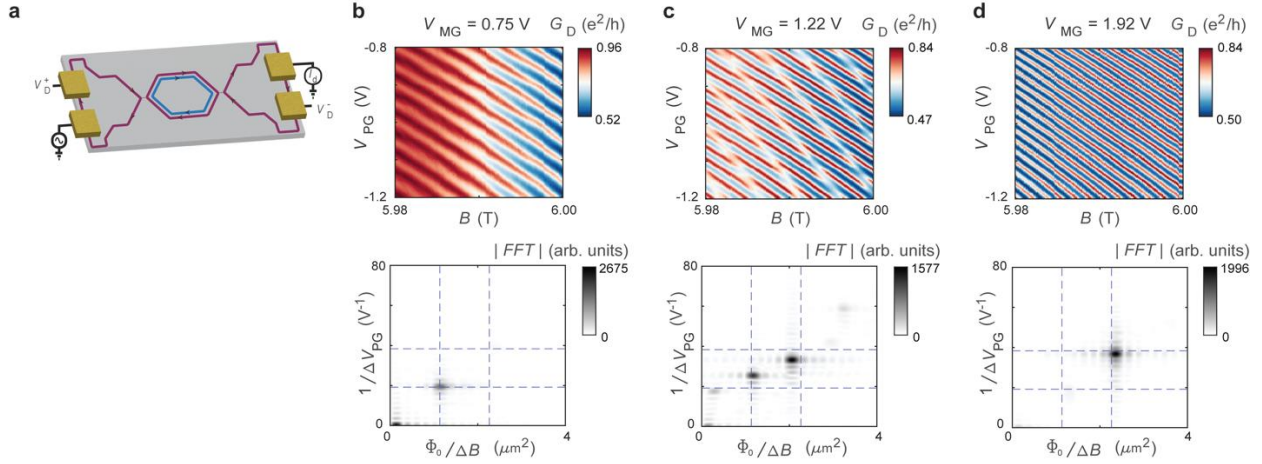

**Fig. S4. Interference of a single edge into a dual-edge cavity. (a)** Schematic of device where left/right regions tuned to $\nu = 1$ allow a lone edge to partially transmit into a cavity at $\nu = 2$. **(b-d)** Diagonal conductance of the outer EC through device at increasing middle gate $V_{MG}$ voltages with corresponding 2D FFTs below.

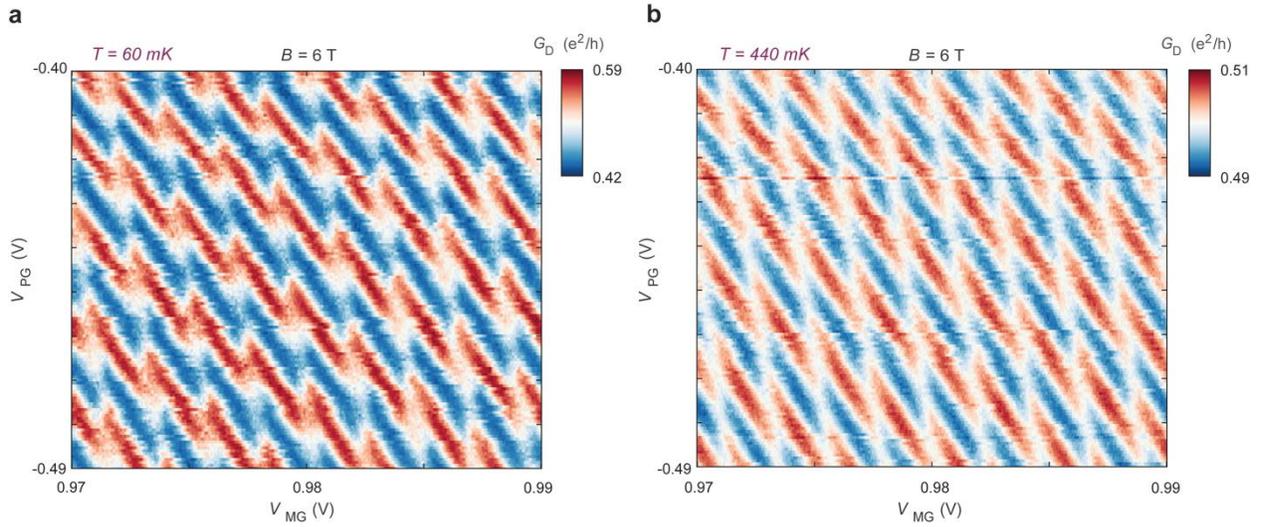

**Fig. S5. Temperature dependence of intermediate-coupling regime.** Conductance oscillations on the outer EC shown at identical sweeping parameters at estimated 60mK **(a)** and 440mK **(b)** electron temperatures.

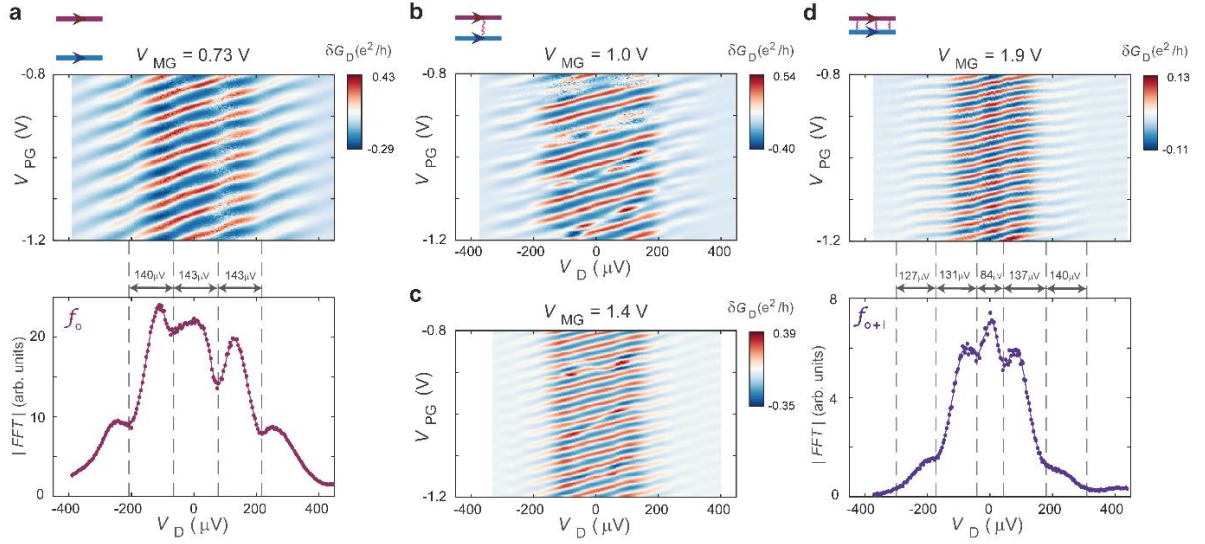

**Fig. S6 | Finite-bias dependence of outer EC interference. a,** $G_D$ oscillations with $V_{PG}$ and $V_D$ in the uncoupled limit. $V_D$ is the measured DC voltage drop on the same probes measuring the AC conductance $G_D$, to get the accurate voltage drop across the FP cavity on the sample for each DC bias applied to the source. Bottom: extracted 1D FFT amplitude at the frequency of the oscillations, $f_o$, as a function of $V_D$ showing the visibility nodes of the interference at finite bias. **b-c,** $G_D$ oscillations with $V_{PG}$ and $V_D$ in two intermediate coupling regime tunings. **d,** $G_D$ oscillations with $V_{PG}$ and $V_D$ in the strongly coupled limit. Bottom: extracted 1D FFT amplitude at the new dominant frequency of the oscillations, $f_{o+i}$, as a function of $V_D$ showing the visibility nodes of the interference at finite bias.